  \magnification=\magstep1
 \settabs 18 \columns
\hsize=16truecm

 \input epsf

\def\b{\bigskip}
\def\bb{\bigskip\bigskip}

\def\no{\noindent}
\def\r{\rightline}
\def\ce{\centerline}
\def\ve{\vfill\eject}

\def\r{\rightline}

\def\harr#1#2{\smash{\mathop{\hbox to .25 in{\rightarrowfill}}
  \limits^{\scriptstyle#1}_{\scriptstyle#2}}}

\def\today{\ifcase\month\or January\or February\or March\or April\or
May\or June\or July\or
August\or September\or October\or November\or  December\fi
\space\number\day, \number\year }

\r \today

\bb\bb\bb




\def\e{\rm e}

\def\p{\partial}

\def\sqr#1#2{{\vcenter{\vbox{\hrule height.#2pt
\hbox{\vrule width.#2pt height#2pt \kern#2pt
\vrule width.#2pt}
\hrule height.#2pt}}}}

  \def\1/2{{\scriptstyle{1\over 2}}}
  \def\a/2{{\scriptstyle{3\over 2}}}
  \def\5/2{{\scriptstyle{5\over 2}}}
  \def\7/2{{\scriptstyle{7\over 2}}}
  \def\3/4{{\scriptstyle{3\over 4}}}

\font\steptwo=cmb10 scaled\magstep2


\def\picture #1 by #2 (#3){
  \vbox to #2{
    \hrule width #1 height 0pt depth 0pt
    \vfill
    \special{picture #3} 
    }
  }

\def\scaledpicture #1 by #2 (#3 scaled #4){{
  \dimen0=#1 \dimen1=#2
  \divide\dimen0 by 1000 \multiply\dimen0 by #4
  \divide\dimen1 by 1000 \multiply\dimen1 by #4
  \picture \dimen0 by \dimen1 (#3 scaled #4)}
  }

%
%

\def\sqr#1#2{{\vcenter{\vbox{\hrule height.#2pt
\hbox{\vrule width.#2pt height#2pt \kern#2pt
\vrule width.#2pt}
\hrule height.#2pt}}}}

\def \r{\rightarrow}

\b

\vskip-2
cm

\ce{\steptwo An equation of state for dark matter}

\b
\ce{C. Fr\o nsdal and T. J. Wilcox}
\b
\ce{\it Department of Physics and Astronomy, University of California Los Angeles}
\b
\no{\it ABSTRACT}  ~Dark matter,  believed to be present in many galaxies,
is interpreted as a hydrodynamical system in interaction with the gravitational field and nothing else. An equation of state determines the mass distribution and the associated gravitational field. Conversely, the  gravitational field can be inferred from observation of orbital velocities of stars in the Milky Way,
in a first approximation in which the field is mainly due to the distribution of dark matter.
In this approximation, the equation of state is determined by  the  gravitational field via the equations of motion.
   
  The potential  is an exact solution of the equations of motion in the approximation of weak fields,
  $$
 - \Delta \phi \propto \rho, ~~ \phi \propto {df\over d\rho},
 $$
 where $f$ is the free energy density. The second equation (the integrated hydrostatic condition) determines $\rho$ in terms of $\phi$; the first equation then becomes a nonlinear equation of the Emden type that issolved exactly by the chosen potential.
 
  The resulting equation of state is a simple expression that accounts for the main features of the  galactic rotation curve over 6 orders of magnitude.   
 
   \b

\no{\bf  I. Introduction}

One of the enduring problems of astrophysics is to place an upper limit 
on the mass of a star. Let us agree from the outset that the  observed mass of a spherically symmetric object is the asymptotic value of the function  $M$ that appears in the  quasi-Schwartzschild metric,
$$
g_{00}(r) = c^2(1-{2M(r)G\over r}).\eqno(1.1)
$$
A locally observed mass is defined in a region where this function is slowly varying.
The Great Attractor near the center of the Milky Way is observed, at a distance from the center of around $ 10^{16}$ cm, to have a value
for this parameter that is several million solar masses  (Ghez 2008). This is as much as 5 orders of magnitude greater than ``reasonable" physical models (Hartle 1978) .

 Analysis of the distribution of velocities of orbiting stars show that   the newtonian potential cannot be attributed to visible sources;  the locally observed mass increases far too rapidly with the distance from the center.  
 Both problems can be qualitatively explained in terms of  `dark matter',  the high
 value of $M$ because the equation of state of dark matter is unknown and not subject to the physical constraints of known forms of matter, the unexpected variation of $M$ with
 distance because dark matter may be present in regions that appear to be empty.

All  that is known about dark matter is that it does not interact with ordinary matter. It does not interact with electrodynamics, it is not in thermal equilibrium with ordinary matter or with radiation,  and the temperature is not
defined.  This puts the theoretician in the same position as he confronts in hydrodynamics when the temperature is eliminated from the theory by means of the
ideal gas equation and the equation of state reduces to  a relation between
density and pressure. The free energy density is a function of density alone, the entropy density  $\p f/\p T$  is zero and the pressure is
$$
p = \big(\rho{\p\over \p \rho}-1\big)f(\rho).\eqno(1.2)
$$
The system is thus determined by the expression chosen for the function $f(\rho)$;   this expression, or the inferred relation between pressure and density,  will be referred to as the equation of state.

In this paper we shall propose a simple, analytic expression for the newtonian
potential that  accounts for  the main features of the rotation curves of
our Galaxy. From this we shall determine the unique equation of state that is required in order that Einstein's equations admit this idealized potential in the   weak field
approximation. Then we use the   equation of state so determined in the full system of
Einstein's equations in the presence of dark matter.

If this  equation of state turns out to be applicable in other galaxies as well,  then this approach to the problem of
dark matter can be considered as an alternative to modified gravity; see for example  Delbourgo (2008) and Mannheim (2011).

\b
{\it Some data.} The radius of the Milky Way is about $r_0 = 10^{23} cm$ and the mass is about $2MG= 2\times 10^{17} cm$. The innermost, observed satellite has a 
nearest approach of  about $2\times 10^{15}cm$ and it moves in  the newtonian field of a mass of about  $2MG = 10^{12}$. The cgs system is used throughout; $1 kpc = 3\times 10^{21} cm$ and  $8\pi G = 1.863\times 10^{-27} cm/g$.
\b

\ce{\it Summary}

Our model for (the negative of) the gravitational potential is
$$
{2MG\over r} =\phi(r)  = k \ln{r+b\over  r},~~ b = \e^{52} cm.
$$
The equations of motion include the hydrostatic condition in integrated form,
$$
{c^2\over 2}\phi  = {d f\over d\rho},
$$
where $f$ is the free energy  function. Einstein's equations, in the weak field approximation, give  a  unique equation of state represented parameterically as follows
$$
f(\rho)= B\psi\sinh^4\psi-p,~~\rho= A \sinh^4{\psi},~~ p =B\int \sinh^4\psi \,d\psi,
$$
 $A$ and $B$  constants.   We solve the relativistic  equations of motion using this equation of state to obtain the gravitational metric and the density distribution of the Galaxy.
\ve
 
\b\b\no{\bf II. The equations of motion}

We shall calculate static, spherically symmetric  solutions of Einstein's equations,
$$
G_{\mu\nu} = {8\pi G\over c^2} T_{\mu\nu},~~ G = .7414\times 10^{-28} {cm\over g},
$$
with a metric of the form
$$
ds^2 = \e^\nu (cdt)^2 - \e^\lambda dr^2 - r^2d\Omega,~~
g_{00}= c^2 \e^{\nu(r)},~~ g_{rr} = -\e^{\lambda(r)},
$$
and a matter energy momentum tensor of the form
$$
T_{00} = \rho U_0U_0,~~ T_{rr} = pg_{00}, 
\eqno(2.1)
$$
all other components zero. Besides Einstein's equation we invoke the hydrostatic
condition in integrated form\footnote {$^1$}{See Section VI.} 
$$
{c^2 \over 2}(\e^{-\nu} -1) = {\p f\over \p \rho}.\eqno(2.2)
$$
 
\b

 The reduced form of Einstein's equations given in the textbooks, beginning with that of Tolman (1934),  is {$^1$} 
$$\eqalign{&
\hskip1cmG_t^t =  -e^{-\lambda}\Big({-\lambda'\over r}+ {1\over r^2}\Big) +
{1\over r^2} =  8\pi G \Big(\e^{-\nu}\rho -p/c^2\Big),\cr 
&\hskip1cm {G_r}^r =  -{\rm e}^{-\lambda }\Big({\nu'\over r} + {1\over r^2}\Big)+
{1\over r^2} =  - 8\pi G   p/c^2.\cr
\cr}\eqno(2.3) 
$$
With the notation \footnote{$^2$}{The choice of the letter $m$ in the first expression is traditional, but unfortunate, in as much the locally observed mass 
defined in (1.1) is $2M(r)G = m(r) + u(r)$. (See below.)}
$$
H(r)=\e^{-\lambda} = 1 -{ m(r)\over r},~ K(r) = \e^{\nu +\lambda} = 1 -{u(r)\over r},
$$
they are
$$\eqalign{&
H' = {1-H\over r} - 8\pi G r  \rho\big(\e^{-\nu} - p/c^2\rho\big),\cr
&
K' =8\pi G H^{-2}  r \rho.\cr}
$$

Some computer programs do not like very large numbers and work better if we change variables, introducing $x$ by 
$$
r = \e^{x},
$$
to get
 $$\eqalign{
 {d \over dx}m(x) &= w\, r^3 \rho\, (\e^{-\nu}-p/\rho c^2), \cr
 {d K(x) \over dx}&= {w\over H^2} \,r^2\,\rho,~~ w := 8\pi G.\cr}\eqno(2.4)
 $$ 
  
 To continue we need an expression for the free energy that will allow us to express
 the density and the pressure in terms of the fields, with the help of Eq.(2.2). To determine the free energy we shall work, provisionally, with the weak  field approximation. 
\b\b

\no {\bf III.  The equation of state} 

A weak field approximation will be used to determine an approximate equation of state, subject to later adjustment. In this approximation we replace Eq.s (2.4) by
 $$
 m'(r) = \, r^2w \rho\,\eqno(3.1)
 $$
 $$
 K'(r) =  \,r\,w\rho.\eqno(3.2)
 $$
 The primes, as before, denote the derivative with respect to $r$.  The   newtonian potential is $-\phi/2$, where
 $$
 1-\e^{-\nu}\approx (1-H) + (1-K) = {m \over r} + {u \over r} = :\phi.
 $$
 Combining (3.1-2) we get
 $$
 m= -r^2 \phi',~~ w\rho=m'/r^2
 $$ 
  The full set of equations is thus
$$
{c^2\over 2}\phi=  {\p f\over \p \rho},~~~  -r^{-2}(r^2 \phi')'= w\rho.  \eqno(3.3)
$$

Something is known about  $\phi$, from observation of radial acceleration of orbiting stars. A family of  satellites moving in circular   orbits with radius $r$ in a  radial,
newtonian  potential  $V$ have  orbital speed $v$ given by $v^2 = rV'$.  Observation has 
revealed that there is a wide interval in which the speed is nearly constant, independent of the distance, which implies that, in this interval, the potential is
approximated by $V = -\phi/2=(k/2)\ln(r)$. We shall model the function $\phi(r)$, then 
calculate the equation of state.  In other words, when the distribution $\phi(r)$ is known from observation, then the last pair of equations provides a parametric representation of the relation between the free energy  density $f$ and the density $\rho$.
Finally  we shall use this equation of state in the exact, relativistic 
field equations. 
\ve
\ce {\it Example} 

Taking
$$
\phi =k \ln{r + b \over r},~~ r = \e^x,\eqno(3.4)
$$
we obtain the velocity distribution shown in  Fig.1  with $k=1$,   $b = \e^{52}$.
 It is very nearly constant for $x<52$ and very nearly newtonian for $ x>> 52$. 
$$
m=-r^2\phi' = kb{ r\over r+b} ~~(= kb - {kb\over r} + ...,~~ r>b),
$$
and the density
$$
w\rho(r) = -r^{-2}(r^2\phi')'= {k\over r^2(1+r/b)^2} .
$$

\vskip3cm

\def\picture #1 by #2 (#3){
  \vbox to #2{
    \hrule width #1 height 0pt depth 0pt
    \vfill
    \special{picture #3}      }
  }
\def\scaledpicture #1 by #2 (#3 scaled #4){{
  \dimen0=#1 \dimen1=#2
  \divide\dimen0 by 1000 \multiply\dimen0 by #4
  \divide\dimen1 by 1000 \multiply\dimen1 by #4
  \picture \dimen0 by \dimen1 (#3 scaled #4)}
  }

\parindent=1pc

\vskip-3cm
 
\epsfxsize.4\vsize
\centerline{\epsfbox{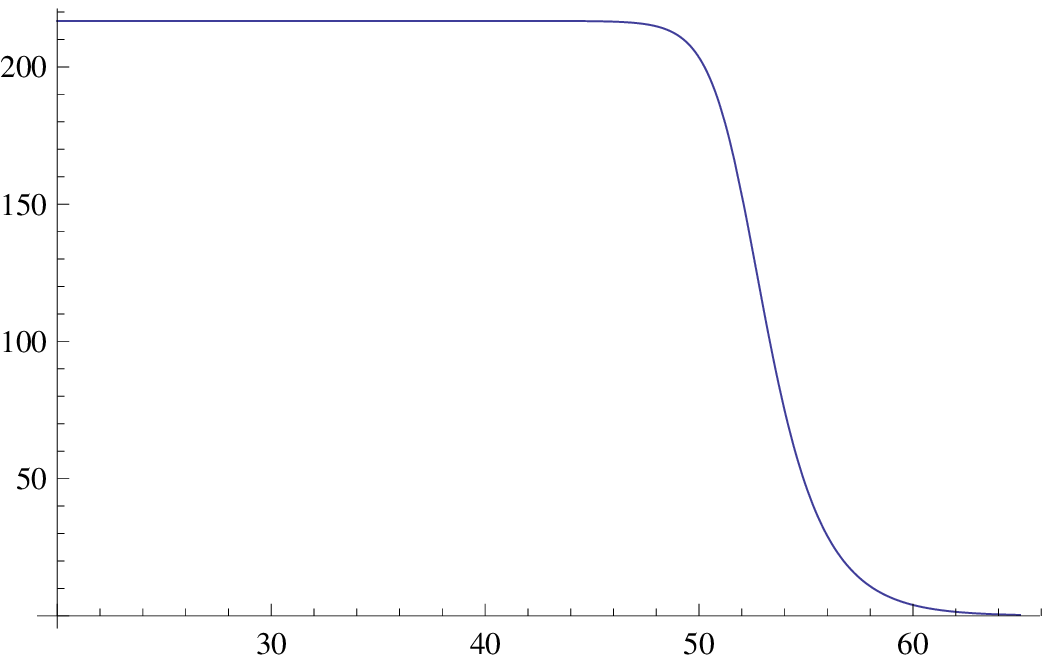}}
\vskip1mm
 
 Fig.1. The orbital velocity distribution  that  was used as a model of the observations in the Milky way.  The velocity is shown in km/sec. 
 \b\b

 The parametric representation of the equation of state is thus, in this case,
$$
 {2\over c^2} {\p f\over \p \rho} = \phi(t) =k\ln{t + b \over t} ,~~w\rho(t) = {k\over t^2(1+t/b)^2}.
$$
Equivalently,
$$
\rho = {16\over b^2}{k\over w}\sinh^4{\psi}, ~~ \psi:= \phi/2k.\eqno(3.5)
$$
The free energy is obtained  by integrating the hydrostatic equation,
 $$
 {2\over c^2}{df \over d \rho} = \phi,~~ {2\over c^2}{df \over d \psi} = \phi{d\rho \over d \psi} =
  2k\Big({d\over d\psi}(\psi\rho) -   \rho\Big).
 $$
 Thus
 $$
 \hat f(\rho)={b^2w\over 16 c^2 k^2 
 }f(\rho) =\int \psi{d\over d \psi}\sinh^4\psi  \, d\psi =\psi \sinh^4\psi-\hat p.
 $$
The last term,
 $$
\hat p := \int\sinh^4\psi\, d\psi = {1\over 32}\Big( \sinh(4\psi)- 8\sinh(2\psi) + 12\psi\Big),
\eqno(3.6)
 $$
 is the pressure,
 $$
p= {16\over b^2}{k^2c^2\over w}\hat p  = \rho{\p\over \p \rho}f - f.\eqno(3.7)
 $$
 It is shown as a function of $\psi$ in Fig.2.
 
 \b

\parindent=1pc
\vskip3cm
 
\vskip-3.cm
 
\epsfxsize.6\hsize
\centerline{\epsfbox{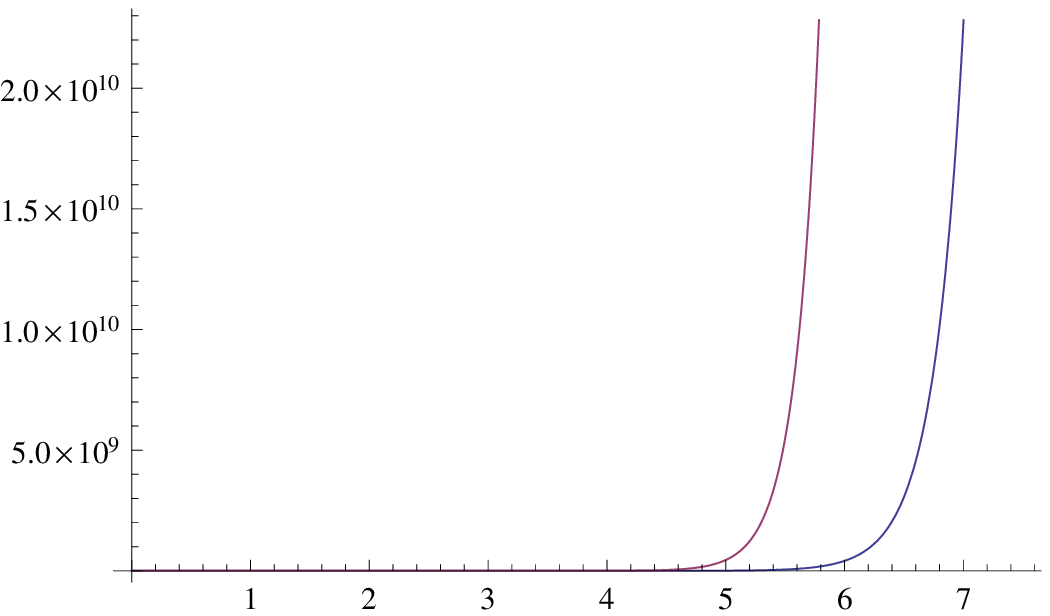}}

Fig.2. Equation of state.  Upper curve: the normalized pressure $\hat p$  as a function of the variable $\psi$,
see Eq.(3.6).  The lower curve
is the $n=4$ polytrope, $\hat p = (1/5)\sinh^5\psi $,  a perfect fit at low values of $\psi$. The innermost 
observed orbiter, at $x = 35$, is at $\psi = 7.5$ and the highest value  (at  $x$ = 21.6)
is 10.5. The deviation from the ordinary polytrope is considerable.
\b\b

Returning to the equations (3.1-2) - the weak field approximation -  we now 
apply the equation of state in the form (3.5-6), fix the appropriate initial values,
$$
x =a=52,~~m(a)={ bk\over 2} = .2\times 10^{17},~~K(a) = 1-k\ln2 +{k\over 2},~~\phi = k\ln 2
$$ 
and run the equations in Mathematica. The program runs from $ x = a$ in both directions, covering 40 orders of magnitude of the radius, correctly reproducing the exact solution (3.4).

\b

{\bf Remark.} The function 
$$
\psi = {1\over 2}\ln(1+b/r)
$$
is an exact solution of the modified Emden equation
$$
\Delta \psi + {8\over b^2} \sinh^4\psi = 0,~~ \Delta = r^{-2}{d\over dr} r^2 {d\over dr}.
$$
The original Emden equation has $\psi^n$ instead of  $\sinh^4\psi$; it has an exact solution in the case that $n = 5$ only.

\ve

\no{\bf IV. Solutions of the relativistic equations of motion}

Using the same equation of state we now solve the exact Einstein equations numerically.     Once the equation of state has been found
  there are no free parameters.

Upgrading the equations, from the weak field approximation (3.3) to the  exact   equations  of motion (2.9) has limited effect, for the fields are relatively weak everywhere;
that is, $\phi << 1$.

After adjustment of the boundary values we  obtained a solution covering the range 
$$
16<x<77,~~~8.9 \times 10^{6} < r<2.8\times 10^{33}.
$$
The values of $m(r)$ at some chosen values of the local mass, are
$$\eqalign{&
m=10^{11}, \hskip.85cm x =39.1,~~  r=9.6\times 10^{16},\cr&
m=   10^{12}, \hskip.9cm x=41.4, ~~r = 9.5\times 10^{17},\cr &
m= 10^{13},\hskip.9cm  x=43.7,  ~~ r= 9.5\times 10^{18},\cr&
m=.4\times 10^{17}, ~~\, x= 58, \,~~r = 1.5 \times 10^{25}.\cr }
$$

The zero of the function $m(r)$ is a computational error. It was verified that the calculation gives result of high accuracy for $ 25<x<60$. 
However, it is not $m(r)$ that should be interpreted as the local mass, but $r \phi(r)$,
since $\phi$ rather than $m/r$  is the newtonian potential. For $ r \phi$ the corresponding values are
$$\eqalign{&
r\phi =  3\times 10^5,\hskip.5cm   x = 23.0,\hskip.4cm r = 9.7\times 10^{9}\cr &
r\phi=10^{11}, \hskip.9cm x =36.4,~~  r=6.4\times 10^{15},\cr&
r\phi=   10^{12}, \hskip.9cm x=38.8, \,\,~\,r = 7.1\times 10^{16},\cr &
r\phi= 10^{13},\hskip.9cm  x=41.3,  \,~~ r=8.6\times 10^{17},\cr&
r\phi=.4\times 10^{17}, ~~ x= 58.0, \,~~r =\,.5\times 10^{25}.\cr }
$$

Observation of the innermost satellites of the Milky Way suggests a local mass of about
$ 10^{12} $ (3 million solar masses) at  a distance of $10^{16}$ from the center.

Other numerical results are as follows. The density is positive; there is a characteristic bump in the density 
profile  - see Fig.3b - where the density reaches the highest value, $ \rho =  100
g/cm^3$ at $x = 21.6$,
$$
\rho_{\rm max} = 100g/cm^3 ~~{\rm at}~~ r =   2.4\times 10^{9}.
$$
This ``object" is comparable to our Sun, in size, mass and gravitational field strength.
  At the shortest distance observed for an orbiting satellite, 
$r = 2\times10^{15} ~ (x = 35.23)$, the density is about $ 1.3\times10^{-10}$. The
pressure has a similar profile - Fig.4, with a peak value of  $2.5\times 10^{16}$.  The  gravitational field $-\phi(x) = c^{-2}g_{00}-1$ also has a maximum  -see Fig.5, reaching a maximum value of $\phi = .00003 $ at the same point. This is about  ten times stronger than the gravitational potential at the surface of the Sun. 
 The appearance of such  shapes is very common when 
polytropic equations of state are used; they are relativistic features not seen in the weak field approximation. For some stars the maximum value
of $\phi$ can rise to get very close to the limiting value of unity, at which point a horizon would appear. 

The local mass predicted by the model at the distance of the inner orbiters  is less than what is observed, by about one order of magnitude.  If this discrepancy can be removed by refinements of the model, then we will have a picture of the galactic
center that is very different from a Schwartzschild black hole.

\vskip3.5cm

\parindent=1pc

\vskip-3cm
 
\epsfxsize.6\hsize
\centerline{\epsfbox{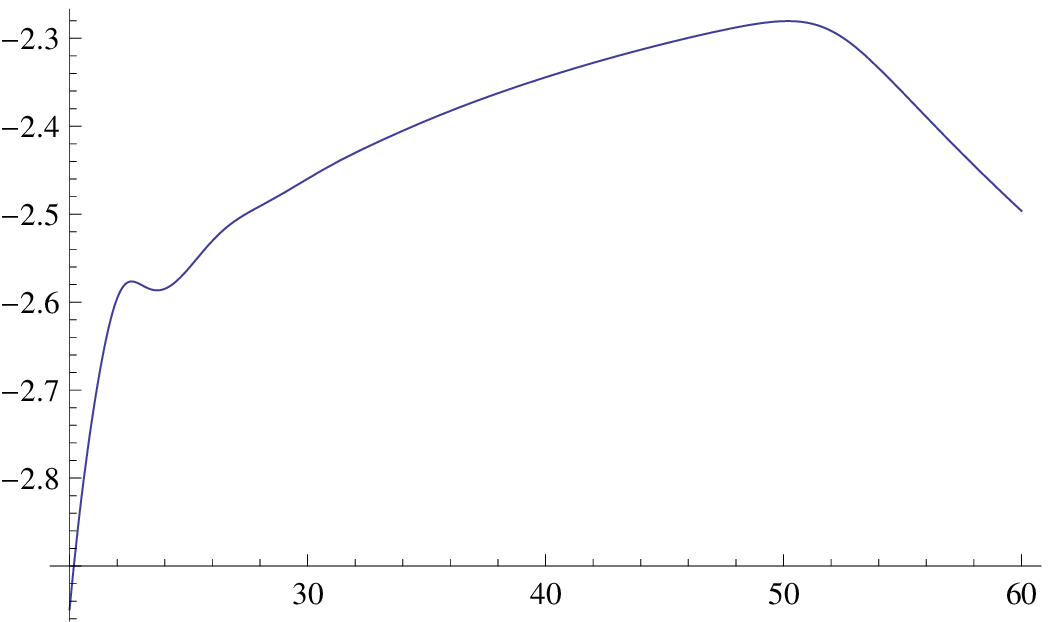}}
\vskip-.1cm

Fig.3. A plot of $\ln\rho/ \ln  r$ against $ x = \ln r$.
 \vskip2cm
\vskip2cm

\def\picture #1 by #2 (#3){
  \vbox to #2{
    \hrule width #1 height 0pt depth 0pt
    \vfill
    \special{picture #3}      }
  }
\def\scaledpicture #1 by #2 (#3 scaled #4){{
  \dimen0=#1 \dimen1=#2
  \divide\dimen0 by 1000 \multiply\dimen0 by #4
  \divide\dimen1 by 1000 \multiply\dimen1 by #4
  \picture \dimen0 by \dimen1 (#3 scaled #4)}
  }

\parindent=1pc   

\vskip-3cm
 
\epsfxsize.6\hsize
\centerline{\epsfbox{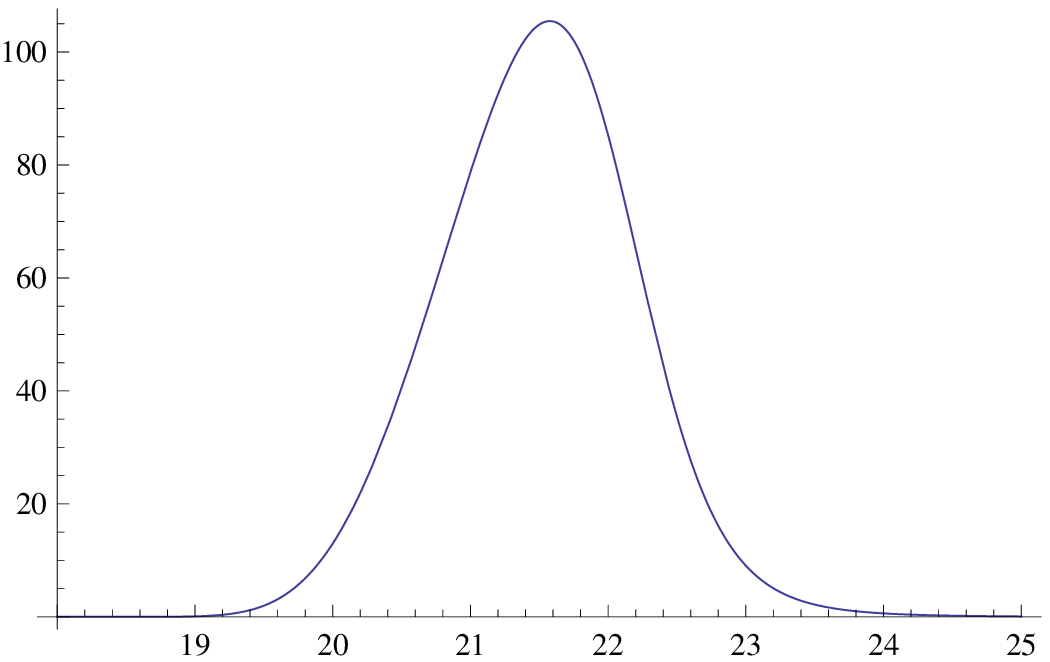}}
\vskip0cm

Fig.4. The characteristic, inner  density profile plotted against $x = \ln r$, with the peak at $r = 9.67 \times 10^{12} cm$.
\ve
\vskip3.5cm

\def\picture #1 by #2 (#3){
  \vbox to #2{
    \hrule width #1 height 0pt depth 0pt
    \vfill
    \special{picture #3}      }
  }
\def\scaledpicture #1 by #2 (#3 scaled #4){{
  \dimen0=#1 \dimen1=#2
  \divide\dimen0 by 1000 \multiply\dimen0 by #4
  \divide\dimen1 by 1000 \multiply\dimen1 by #4
  \picture \dimen0 by \dimen1 (#3 scaled #4)}
  }

\parindent=1pc   

\vskip-3.2cm
 
\epsfxsize.6\hsize
\centerline{\epsfbox{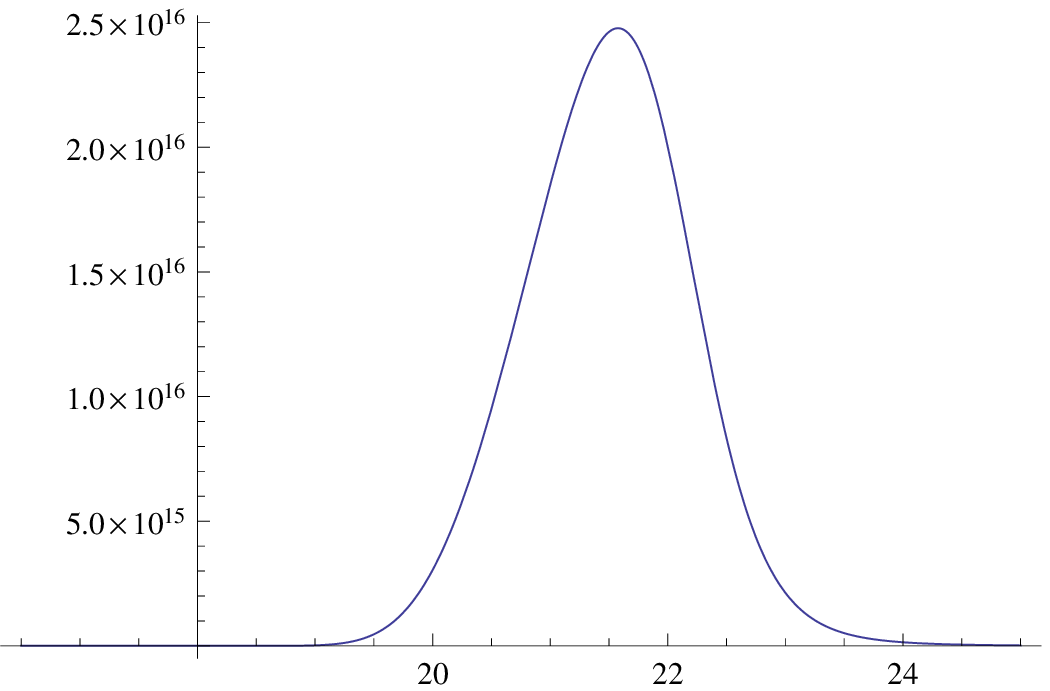}}
\vskip0cm

Fig.5. The pressure profile in the inner region..

\vskip3.5cm

\def\picture #1 by #2 (#3){
  \vbox to #2{
    \hrule width #1 height 0pt depth 0pt
    \vfill
    \special{picture #3}      }
  }
\def\scaledpicture #1 by #2 (#3 scaled #4){{
  \dimen0=#1 \dimen1=#2
  \divide\dimen0 by 1000 \multiply\dimen0 by #4
  \divide\dimen1 by 1000 \multiply\dimen1 by #4
  \picture \dimen0 by \dimen1 (#3 scaled #4)}
  }

\parindent=1pc

\vskip-3cm
 
\epsfxsize.6\hsize
\centerline{\epsfbox{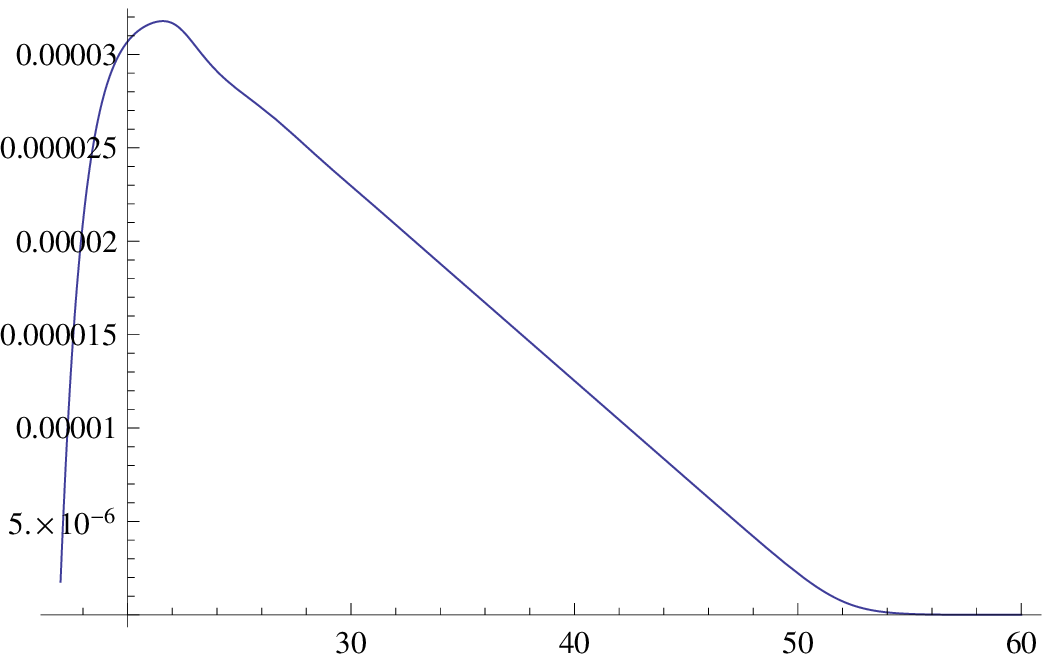}}
\vskip0cm

Fig.6. The potential has a maximum at the same point. The density and pressure peaks are narrower because of the high value of the ``polytropic index".

\b
\b

\parindent=1pc
\vskip3cm
 
\vskip-3.cm
 
\epsfxsize.6\hsize
\centerline{\epsfbox{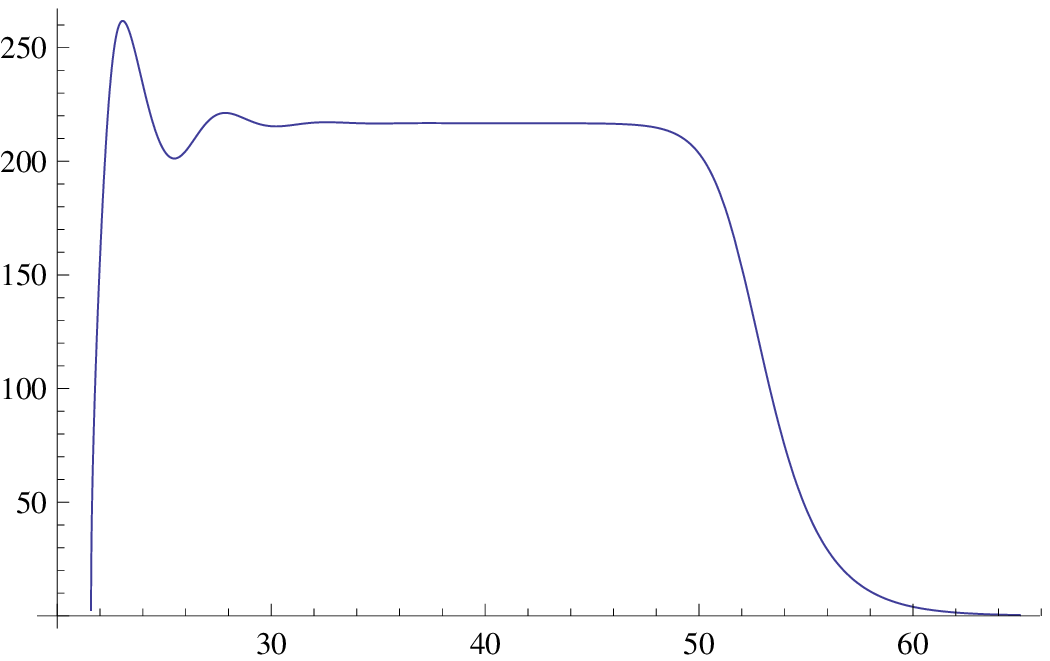}}

Fig. 7. The velocity distribution predicted by the relativistic equations of motion, in km/sec.   Compare Fig.1. 
 
\ve
\no{\bf V. The nature of dark matter}

The equation of state was obtained in parametric form, Eq.s (3.5) and (3.6),
$$
  \rho =C \sinh^4\psi,~~p = {kc^2\over 32}C(\sinh4\psi - 8 \sinh 2\psi + 12\psi ),\eqno(5.1)
$$
with
$ 
C = 16 k/b^2 \omega.
$
It bears a remarkable   similarity to an equation first proposed by Stoner(1932) and  used by Chandrasekhar (1935),\footnote {$^3$}{It was `corrected'  and used by Oppenheimer and Volkov  (1939)  in their study of  neutron stars.
The original  version, quoted here,  is in Landau and Lifshitz (1958) page 168.} 
$$
\rho = C_1\sinh^3t,~~ p =C_2 \big(\sinh 4t - 8 \sinh 2t + 12 t \big).
$$
The basis for this formula is a model of fermions in a collapsed state, the Femi sea being filled up to $q/m = \sinh\psi$.

The similarity, if not regarded as a coincidence, suggests that dark matter  may be a cloud of ``ice crystals" that consist of fermions in a highly reduced state.
The total absence of interactions, and of photons, is a premise of Chandrasekhar's work.
  A slightly different model reproduces our equations (5.1) exactly,
but the fermions need to have an additional degree of freedom; for example, an extra dimension of momentum space. With $q = \sinh\psi, E = \cosh\psi$, integrating over the 3-sphere with radius $q$,\footnote{$^4$}{  The factor $E$ in the first integral arises because the mass of the compound system is the sum of the energies of the constituents. This factor cancels 
the factor $1/E$ in the volume element, just as in the calculations of Stoner and Chandrasekhar. }
$$
\hat\rho =  q^4 ={1\over2\pi^2}\int E \,{d^4q\over E},~~ \hat p =\int q^4d\psi =
{1\over 2\pi^2}\int q \,{ d^4q\over E}.\eqno(5.2)
$$
 The equation of state   that  has been developed here is consistent with the observed velocity  distribution, even for the innermost  orbiters, but  this does not give enough information to develop a microscopic model of dark matter.

It is essentially a   hydrodynamical system. At very low densities the star is an $ n = 4$
polytrope, 
$$
 \hat p = {1\over5}\, \hat \rho^{5/4}, ~~ \hat \rho = \sinh^4\psi.
 $$
 With 
 $$
 \rho =\alpha \hat\rho, ~~\alpha = {16 k\over b^2 w} = 6.1 \times 10^{-24},
 $$
  and 
 $$
 p =\alpha\beta \hat p,~~ \beta =  c^2 k  = 9.4 \times 10^{14},
 $$
 it works out to
 $$
 p = A \rho^{5/4}, ~~ A = {1\over 5}\alpha^{-1/4}\beta = 6.0 \times 10^{20}.
 $$
 At very high densities the adiabatic index is effectively   infinite, $\hat p = \hat\rho/4$ and
 $$
  p = {\beta\over 4} \rho = 2.35\times 10^{14}\rho.
$$ 
 This relation is confirmed at  the point where $\psi = 10.48$,    
 $\rho = .00122 g/cm $, where we are in the regime of high densities.

\b\b

\no{\bf VI. Theory}

The reader will have noticed that the equations that were used are classical,
but a closer look reveals some novelties.
\b
1. The integrated hydrostatic condition is classical; a short calculation shows that
taking the  gradients of both sides leads to the usual hydrostatic condition,
$$
\rho\,{\rm grad}\, \phi = -{\rm grad \,p}.
$$
The difference is that the integrated form incorporates the boundary condition that
fixes the speed of light at infinity.     

2. It is of interest to ask why this boundary condition has not been applied previously.
One part of the answer is that  the stars of Eddington and Chandrasekhar  all have abrupt boundaries at a point where the temperature and the density vanish, so that these fields are not continuous.
Another part of the reason is that the equations employed by Eddington (1926),  
Chandrasekhar (1935) and many others
differ in one particular from ours: The factor $\e^{-\nu}$ that multiples the density
in Eq.(2.8) is absent; consequently, the function $\nu$  is represented only by its derivative,  so that
fixing its boundary value has no meaning. 

In view of this difference we need to justify our approach. All equations used are variational equations based on the following action,
$$
A = {1\over 8\pi G}\int d^4x \sqrt{-g} R +
 \int d^4x \sqrt{-g}\Big(\rho(g^{\mu\nu}\Psi_{,\mu}\Psi_{,\nu} - c^2) - f(\rho)\Big).
 $$
The non relativistic approximation of the matter lagrangian is one that was used
by Fetter and Walecka (1980) to obtain a variational formulation of hydrodynamics: the equation of continuity and the Bernoulli equation (in integrated form).\footnote{$^5$}{The non relativistic approximation is taken by setting $\Psi = c^2 t + \Phi$, neglecting $1/c^2$
terms and interpreting $\Phi$ as the velocity potential.}  
The term $f(\rho)$ is the  free energy, as is seen from the structure of the equations of motion
(Fr\o nsdal 2007, 2008). The variational approach to hydrodynamics is an application of the Gibbs variational principle to the case that the temperature is frozen, so that neither temperature nor entropy plays any role, as is appropriate for a treatment of dark matter,
effectively a hydrodynamical system. The density $\rho$ is denoted $\rho + p$
by Eddington; this is a matter of notation, and irrelevant in the 
immediate context, since $p/c^2 \rho$  never exceeds $10^{-4}$. The 
gradient of the field $\Psi$ corresponds to Tolman's vector field $U$; for a stationary solution $\Psi_{,0}$ is a constant, while Tolman's normalization condition leads to
$U_0 = \sqrt{g_{00}}$. This is what gives rise to the cancellation of this metric function
in Tolman's equations of motion, and it constitutes an important difference in principle between our approach and that of Tolman.\footnote{$^6$}{ In a variational approach
it is important to specify the independent variables; any constraint  is the source of great complications.}

3. Another consequence of action principle dynamics is that the current is conserved,
$$
\p_\mu J^\mu = 0,~~ J^\mu =  g^{\mu\nu}\Psi_{,\nu} \rho.
$$
The usual approach   does not admit a conserved current
and breaks with non relativistic theory in this respect.

Although the current is conserved, it is not directly related to the ``mass" as we would define it. Our approach  to stellar structure is to start the analysis from the outside,
using observational data. In the models considered here the metric has the asymptotic form
$$
c^{-2}g_{00} = 1 - {2MG\over r},
$$
with $M$ constant. This is an obervational datum, measured by observing the motion of test bodies. In this paper, that is what we call the  mass of the star.  Now it is always pointed out that  the function $m$ that appears in (2.4) tends to $M$ at infinity.
 In the traditional approach  the equation is just $dm/dr = wr^2
 \rho$ and 
 the mass can be expressed as an integral over the density, provided that $m$ vanishes at the origin.    \footnote{$^7$}{It has   
been pointed out that the integral $\int\rho d^3 r$ does not have the correct measure. Kippenhahn and Weigert (1990) calls this situation hazardous, but no one seems to have taken the warning seriously.}  
Our theory preserves the continuity equation of classical hydrodynamics, but there is no direct connection between mass and the conserved quantity.  \footnote{$^8$}{It is true that the increment of mass, between distances $r_1$ and $r_2$ from the center, is the integral of the density  over the region bounded by two spheres. Note, nevertheless, that the non relativistic gravitational potential
 arises entirely from the time component of the metric, while the function $m$ is in the space component. The unfortunate association of this function with the mass is due to Tolman's normalization condition.}  Consequently, there is no need to postulate that the function $m$ vanishes at the origin.

\b
 An equation of state in hydrodynamics  is a relation between density and pressure.
The pressure term in Einstein's equation is not as important as the role that is played by the pressure in the hydrostatic equation.  In our approach  the equation of state follows from the expression that is chosen  for the free energy density. Our approach preserves all the structure of hydrodynamics, including   the equation of continuity.
\b
Ultimately, the model of our Galaxy must be improved by including the 
contribution of visible matter;  the overall structure can be studied 
in terms of an idealized, continuous, spherically symmetric distribution
that makes an important, additional  contribution to the action.
The absence of any interaction between dark and visible matter makes this
very straightforward; in the absence of any interaction between the two kinds of matter the free energy density is additive. Note that it is
essential to recognize the roles of two quite different  density fields.

\b\b
\no{\bf VII. Speculating about the center}

In this paper the study of our galaxy has been approached from the outer regions,
because that is where observations have been made up to this time. The analysis is
especially interesting because nothing is known about the nature of dark matter.
Consequently, there is no need to ask what pressures and densities can be
allowed; there is no way that we can answer  questions of this kind.

Historically, a number of statements have been made that would place limits on the mass of certain types of stars. These statements all rely on two assumptions:\break 
(a) that the nature of the matter within the star is within the limits of our knowledge
and (b)  that the total mass is  the integral of the density, from the center outwards.
If these premises may said to be  reasonable as  far as stars made up of ordinary matter is concerned, they cannot be applied with any degree of confidence to the
new situation that is faced in connection with dark matter. The bump in the density
is a common feature of relativistic models; here it is predicted to occur at a distance of about $2.4\times 10^{9} cm$ from the center. It is conceivable that future observation
may validate this prediction, but what happens inside is beyond our reach.

 \b\b
\no{\bf Acknowledgements}

We are grateful to  Robert Delbourgo, Paul Frampton, Philip Mannheim, John Moffat and  Abhishek Pathak for stimulating conversations.
 
\b
\no{\bf References}

\no Chandrasekhar, S.,   Month.Not. R.A.S. {\bf 95} 222 (1935).  
 
\no Chandrasekhar, S., {\it An introduction to stellar structure}, U. Chicago press 1938.
 
\no  Delbourgo, R. and Lashmar, D.   ``Born Reciprocity and the  $1/r$  potential", 

Found. of Phys. 38, 995-1010, 2008.

\no Emden,  J. R., {\it Gaskugeln}, Teubner 1907.

\no Eddington, A.S.,    {\it The internal constitution of stars}, Dover. N.Y., 1959.

\no Fetter, A.L. and  and Walecka, J.D., {\it Theoretical mechanics of particles and continua},  

McGraw-Hill, N.Y. 1980.

 \no Fr\o nsdal,    C. ``Heat and Gravity. I. The action principle",   arXiv:0812.4990. 

 \no Fr\o nsdal,  C.  ``Ideal stars and General Relativity", Gen.Rel.Grav. {\bf 39} 1971-2000 (2007).
 
\no Ghez, A. M., Salim, S., Weinberg, N. N., Lu, J. R., Do, T., Dunn, J. K., Matthews, K., 

Morris,
M., Yelda, S., Becklin, E. E., Kremenek, T., Milosavljevic, M., Naiman, J., 

 ÒMeasuring
Distance and Properties of the Milky WayÕs Central Supermassive Black 
 
 Hole with Stellar
Orbits,Ó ApJ, 689, 1044 (2008).
 
\no Gibbs, J.W., ``On the equilibrium of heterogeneous substances",

Trans.Conn.Acad.  108-248 (1878).

\no Hartle, J.B. ``Bounds on the mass and moment of inertia of non-rotating neutron stars", 

 Physics Reports {\bf 46} 201-247 (1978).

\no Kippenhahn,  R. and Weigert, A., {\it  Stellar structure and evolution}, Springer-Verlag  1990.

\no Landau, L. D. and Lifshitz, E. M., {\it Statistical physics}, Pergamon Press 1958.	 

\no Mannheim, P., ``Making the case for conformal gravity", arXiv 1101.2186.

\no Oppenheimer, J.R. and Volkov, G.M., ``On massive neutron stars", 

Physical Review {\bf 55} 374-381 1939).

\no Stoner, E.C.,  ``The minimum pressure of a degenerate electron gas",  

Monthly Notices R.A.S., {\bf 92} 651-661 (1932).

\no Tolman,  R.C., {\it Thermodynamics and Cosmology}, Clarendon, Oxford 1934.  

  \end